\documentclass[pre,floatfix,twocolumn,noshowpacs,groupedaddress]{revtex4-1}

\usepackage{amsmath,amsfonts,mathrsfs}
\usepackage{epsfig,graphicx}
\usepackage{color}
\usepackage{multirow}
\usepackage[caption=false, font=footnotesize, justification=RaggedRight]{subfig}
\graphicspath{{./Figs/}}

\usepackage{array}
\newcolumntype{H}{@{}>{\lrbox0}l<{\endlrbox}}

\begin{document}

\title{The Nobel Prize delay}
\author{Francesco Becattini}
\affiliation{Universit\'a di Firenze and INFN Sezione di Firenze, Florence, Italy} 
\author{Arnab Chatterjee}
\author{Santo Fortunato}
\affiliation{Department of Biomedical Engineering and Computational
  Science, Aalto University School of Science, P.O.  Box 12200,
  FI-00076, Finland} 
\author{Marija Mitrovi\'c}
\affiliation{Scientific Computing Laboratory, Institute of Physics Belgrade, University of Belgrade, Pregrevica 118, 11080 Belgrade, Serbia}
\author{Raj Kumar Pan}
\author{Pietro Della Briotta Parolo}

\affiliation{Department of Biomedical Engineering and Computational
  Science, Aalto University School of Science, P.O.  Box 12200,
  FI-00076, Finland} 

\begin{abstract}
The time lag between the publication of a Nobel discovery and the
conferment of the prize has been rapidly increasing for all disciplines,
especially for Physics. Does this mean that fundamental science is 
running out of groundbreaking discoveries?

\end{abstract}

\maketitle

The 2013 Nobel Prize in Physics was awarded to Higgs and Englert for their prediction of the existence of the Higgs boson. 
Though the Higgs particle was experimentally discovered at CERN in 2012, the original theoretical works date back to the 1960s. 
Thus, it took about half a century of intense work to confirm their prediction.  

Long time lags between discovery and recognition are not unusual. In fact, it has been significantly increasing over the years (Figure 1).  
\begin{figure}[h]
  \includegraphics[width=0.99\linewidth]{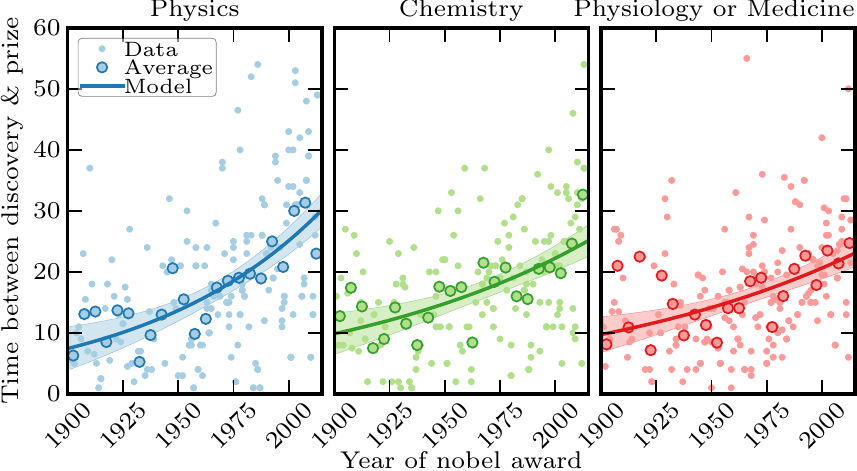}
  \caption{Time difference (in years) between the discovery and the awarding of the Nobel prize, versus the year when the award is received. Each plot shows the raw data, the 5-year average, and the exponential fit with its confidence interval. The lag is increasing for the three fields, with rates of 0.012$\pm$0.002, 0.008$\pm$0.002 and 0.008$\pm$0.001 for Physics, Chemistry and Physiology or Medicine, respectively.}
\label{fig1}
\end{figure}
Let $\Delta^{\mathrm{D \rightarrow N}}$ be the time between the discovery and the Nobel award. We model the variation of $\Delta^{\mathrm{D \rightarrow N}}$ with 
time $t$ by considering an exponential law:
\begin{equation}
  \Delta^{\mathrm{D \rightarrow N}}(t) = c_{\alpha}\exp(\alpha t) ,
  \label{eq:NSinceD_exp}
\end{equation}
where $\alpha$ is the rate of increase in $\Delta^{\mathrm{D \rightarrow N}}$ and $c_{\alpha}$ is a proportionality constant. Figure 1
shows an increase in $\Delta^{\mathrm{D \rightarrow N}}$ for all fields. The predicted values and indicated 95\% confidence intervals 
are given by the exponential regression model. 
In order to validate our regression analysis we check the residuals,
as measured by the difference between the observed and the predicted
values. The linear regression model shows non-random patterns in the
residual plots. 
A piece-wise linear regression shows that the slopes of the curves
consistently increase with time for all the fields. We concluded that
the observed increase is faster than linear.
In order to fit the data using a model with minimum number of
parameters,
consistent with the above observation, we used the exponential
regression model of Eq.~\ref{eq:NSinceD_exp}. Indeed, the exponential
fit turns out to be better than the linear one, although the 
conclusions are essentially the same in both cases.
The rate of increase in $\Delta^{\mathrm{D \rightarrow N}}$ is the highest for Physics, 
followed by Chemistry and by Physiology or Medicine. On the x-axis of Fig. 1 we report the year when the Nobel Prize was actually awarded. This means that
future awards for already published discoveries will have no influence on the ones shown in our plots, they will contribute to the future evolution of the curves.

In order to corroborate our results about the growing time lag between
discovery and prize we consider the frequency of awards within
T years of discovery (20 years in Fig. 2). Using a logistic regression
model we measure whether this probability is decreasing with
time and how.
Simple logistic regression is analogous to linear regression, except
that the dependent variable is nominal, not a measurement.
In the present case it would be ``prize within 20 years of discovery''
or ``prize after 20 years of discovery''. 
The goal is to see whether the probability of getting a particular value of the nominal variable (binary response in our case) is associated with the measurement variable (time).
\begin{figure}[h]
  \includegraphics[width=0.99\linewidth]{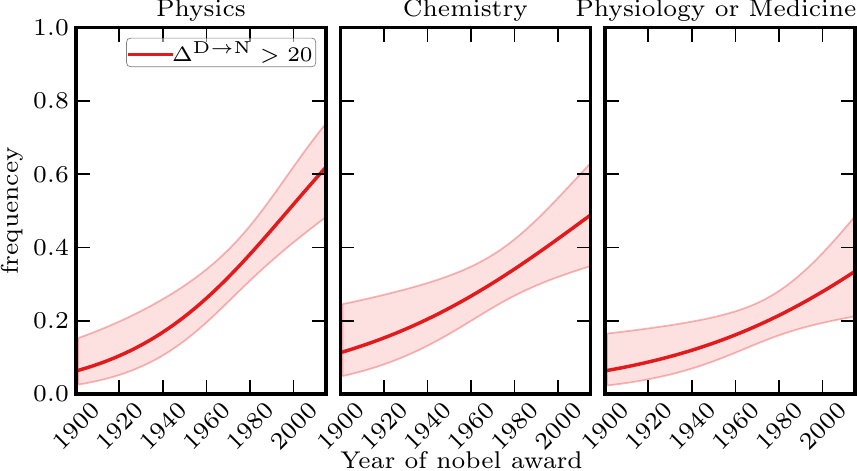}
  \caption{The frequency of prizes awarded over 20 years of the discovery is increasing for all disciplines. The growth is fastest for Physics and slowest for Physiology or Medicine. }
\label{fig2}
\end{figure}
We estimate first-degree logistic polynomial regressions for all
the fields and show the predicted values and 95\% confidence intervals. The conditional probability of the discovery being awarded within $T$ years is given by 
\begin{equation}
  \mathrm{Pr}(\Delta^{\mathrm{D \rightarrow N}}<T|t)=\frac{1}{1+\exp[-(\mu+\nu t)]}, 
  \label{eq:NSinceD_logistic}
\end{equation} 
where the parameters $\mu$ and $\nu$ are estimated using the maximum
likelihood method.
\begin{figure*}[htb]
  \includegraphics[width=\textwidth]{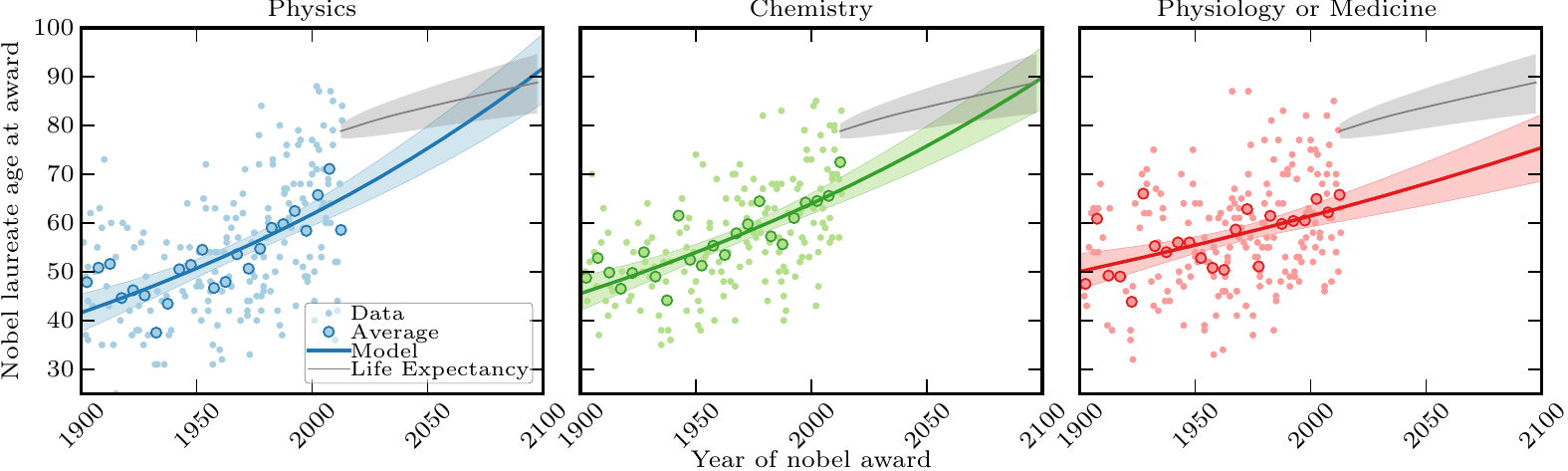}
  \caption{Change in the age of the scientist at which the Nobel prize is awarded. For all fields there is an increasing trend. 
For Physics and Chemistry the rate of increase is similar ($0.0040\pm0.0005$ and $0.0034\pm0.0004$), while for Physiology or Medicine the increase is much smaller ($0.0020\pm0.0005$). 
The progression of the average life expectancy in the United States is also shown in grey.}
\label{fig3}
\end{figure*}
After 1985, about 15\% of Physics,  18\% of Chemistry and 9\% of Physiology or Medicine prizes are awarded within 10 years of their discovery. In contrast, before 1940 about 61\% of Physics, 48\% of Chemistry and 45\% of Physiology or Medicine prizes are awarded within 10 years of the discovery. Correspondingly, after 1985 about 60\% of Physics, 52\% of Chemistry and 49\% of Physiology or Medicine prizes are awarded over 20 years of the discovery. In comparison, before 1940 only about 11\% of Physics, 15\% of Chemistry and 24\% of Physiology or Medicine prizes were awarded over 20 years of the discovery. In all fields the frequency of the prize being awarded over 20 years since discovery is increasing. The rate of increase in the frequency of getting the award 20 plus years since the discovery is fastest for Physics and slowest for Physiology or Medicine.

As a result of the increasing time to recognize a Nobel discovery, the
age at which laureates receive the award is also increasing. This is
also determined by the increasing age at which major discoveries are
made~\cite{jones11}.
We consider how the age at which scientists
are awarded the Nobel prize $a^{\mathrm{N}}$ is changing with time. An exponential increase is represented by  
\begin{equation}
  a^{\mathrm{N}}(t) = c_{\gamma}\exp(\gamma t),
  \label{eq:nAge_exp}
\end{equation}
where $\gamma$ is the rate of increase of the age and $c_{\gamma}$ is a proportionality constant.

Figure 3 shows that  $a^{\mathrm{N}}$  is increasing for the three
fields. We also used the regression model 
to project the age of the laureates at the time of the award until the
end of the century. 
The predicted values and indicated 95\% confidence intervals are given by the exponential regression model. 
The figure also shows the projected life expectancies (of men and women combined together) across the 21st century. 
Here we used the data of the United States as a proxy of the life expectancy (as US citizens have been awarded the majority of Nobel prizes). 
The expectancy is based on WPP2012 estimates using the medium scenario and the 95\% prediction interval is also shown~\cite{UN}. 
We found that by the end of this century for the fields of Physics and Chemistry, the Nobel laureates' age at discovery would become higher than the life expectancy.

Therefore, if this trend is maintained, by the end of this century the average
age of Nobel prize awarding shall reach life expectancy, with easily conceivable
consequences. 

What is the reason of the increasing delay between discovery and
recognition ?
To make the discussion serious, one should first take into account the effects of the 
change of society on scientific research, which may affect Nobel prize awarding. 
The increasing number of scientists, the increasing life expectancy, the changing 
research and career policies, the increasing training time etc. must all certainly 
play a role. Yet, one can reasonably assume that these factors affect all disciplines 
to a comparable extent and cannot explain the remaining differences, like the one
which is observed between Physics and Medicine. Under this assumption, our analysis
suggests that frequency of groundbreaking discoveries in fundamental science is
decreasing. It is well known that fundamental science has undergone major 
changes over the past century, becoming a highly collaborative endeavour requiring 
a sustained effort by large teams over many years and yet, this might just be another 
facet of the problem, namely the increasing objective difficulty in achieving progress.  
An alternative interpretation of the difference between Physics and Medicine 
would be its converse: since no more than two discoveries can be awarded the Nobel 
prize at the same time, it could be that the number of important discoveries is
in fact increasing, and that, in order not to lose worthy winners, one is forced 
to dig deeper and deeper in the past. However, we think that this interpretation 
is disfavoured. First, it should be noted that, on the
average, recognition takes a longer time, but it is also true that, even in modern times, 
there have been major discoveries that have been quickly recognized and awarded 
(e.g. graphene). Secondly, if the frequency of breakthroughs was stable or even 
increasing in time, the fluctuations with respect to the average ``importance'' 
of each awarded discovery would likewise increase in time. Thus, one would expect
a comparable or higher frequency of exceptional advances (relativity-like hits), 
in recent times, which does not seem to have occurred. Indeed, it is hard to imagine 
that such extra-important discoveries would be held in stand-by to recognize
less important discoveries of the past.

\section*{Data}
We collected data on dates of birth, the year of Nobel prizes and year(s) of publication(s) of prize winning work. 
As a primary data source we used the Nobel Foundation's website, \texttt{nobelprize.org}. In the cases where the information was not 
sufficient to accurately identify year(s) of prize winning publication we consulted all the publications of the Nobel Laureates 
using \texttt{google.scholar.com}. We then determined the year of the most relevant  publication related to the topic of the 
Nobel prize award. We also consulted the biographies of the laureates and other resources, 
such as \texttt{nobel.caltech.edu/}, \texttt{journals.aps.org/prl/50years/milestones}.

\end{document}